\documentclass{article}
\usepackage{spconf,amsmath,graphicx}
 \usepackage{booktabs}
 \usepackage{array}
 \usepackage[bookmarks=false]{hyperref}
 \usepackage{xcolor}
 \usepackage{cleveref}
\usepackage{cite}
\usepackage[ruled,vlined]{algorithm2e}
\title{A study on the efficacy of model pre-training in developing neural text-to-speech system }

\name{Guangyan Zhang$^1$, Yichong Leng$^2$, Daxin Tan$^1$, Ying Qin$^3$, Kaitao Song$^4$, Xu Tan$^4$, Sheng Zhao$^5$, Tan Lee$^1$}
\address{
  $^1$ Department of Electronic Engineering, The Chinese University of Hong Kong\\
  $^2$ University of Science and Technology of China\\
  $^3$ Institute of Information Science, Beijing Jiaotong University, Beijing 100044, China\\
  $^4$ Microsoft Research Asia $^5$ Microsoft Azure Speech }

\begin{document}
%\ninept
\maketitle

\begin{abstract}
In the development of neural text-to-speech systems, model pre-training with a large amount of non-target speakers' data is a common approach. However, in terms of ultimately achieved system performance for target speaker(s), the actual benefits of model pre-training are uncertain and unstable, depending very much on the quantity and text content of training data. This study aims to understand better why and how model pre-training can positively contribute to TTS system performance. It is postulated that the pre-training process plays a critical role in learning text-related variation in speech, while further training with the target speaker's data aims to capture the speaker-related variation. Different test sets are created with varying degrees of similarity to target speaker data in terms of text content. Experiments show that leveraging a speaker-independent TTS trained on speech data with diverse text content can improve the target speaker TTS on domain-mismatched text. We also attempt to reduce the amount of pre-training data for a new text domain and improve the data and computational efficiency. It is found that the TTS system could achieve comparable performance when the pre-training data is reduced to 1/8 of its original size.
\end{abstract}
\vspace{-1.5em}
\section{Introduction}
\label{sec:intro}

\vspace{-0.1em}
In recent years, neural text-to-speech technology has demonstrated significant successes in generating high-quality speech with good naturalness and expressiveness for a target speaker \cite{tan2021survey,wang2017tacotron,shen2018natural,elias2021parallel,ren2020fastspeech}. In research and development of neural TTS, the text content of training and test data are often highly similar and in the same text domain. For many real-world applications, TTS systems need to deal with text input with arbitrary content across a wide range of domains. Their performance may deteriorate substantially on domain-mismatched text \cite{he2019robust} due to the limited content and domain coverage of training data. It is generally costly or impractical to increase the quantity and diversity of training data for a specific target speaker, whilst speech data from other ``non-target'' speakers may be easily accessible and available. Leveraging large amount of non-target speakers' data from different sources has become a common and appealing approach to developing high-performance TTS systems when training data from the target speaker(s) are limited \cite{chung2019semi,arik2018neural,chen2018sample,cooper2020zero, tan2021cuhk}. However, it was noticed that the actual benefits of using training speech from other speakers could be uncertain and unstable \cite{chung2019semi}. Understandably the benefits depend very much on the content and quality of non-target speakers' data. There are three aspects of consideration: (1) coverage and domain of text content; (2) speaker similarity with respect to the target speaker; and (3) acoustic condition. The present study mainly focuses on the aspect of text content. 

\vspace{-0.2em}
Our preliminary explorations showed that TTS systems trained solely on target speaker's data did not perform well in predicting appropriate prosody for domain-mismatched input text. Prosody is utmost important in determining the naturalness and expressiveness of speech. We postulate that speech prosody can be seen as the combination of two components of variation in speech, namely text-based variation and speaker-based variation. The text-based component, termed as text prosody henceforth, refers to the general prosodic characteristics that are basic and essential to expressing the intended text content, e.g., lexical tone, lexical stress, sentence intonation \cite{taylor2009text}. If text prosody is not realized or controlled properly, the synthesized speech would sound unnatural and inappropriate, even if the pronunciation is largely correct. In \cite{aoyama2007prosody}, native listeners were found to be capable of detecting abnormality in the non-native speech that has International Phonetic Alphabet (IPA) transcription identical to that produced by native speakers. The speaker-based component of prosodic variation\cite{jia2018Transfer}, termed as speaker prosody, is concerned primarily with an individual's speaking style. In particular, we differentiate it from timbre, which refers mainly to voice (phonation) characteristics.

\vspace{-0.2em}
We conjecture that text prosody can be captured by pre-training the TTS model on a large amount of speech data with diverse text content and involving multiple non-target speakers. A speech generation system fine-tuned from such pre-trained model with target speaker's data is expected to have better performance on domain-mismatched text input than a corresponding system without using pre-trained model. In a typical multi-speaker TTS model design\cite{yamagishi2009robust,yang2016training}, text content and speaker identity are processed by separate modules. It adopts a speaker-independent TTS model with text embeddings and speaker embeddings. This speaker-independent TTS model deals with only text information regardless of speaker variation. After being fine-tuned on target speaker data, the speaker-independent TTS model would supposedly retain the text prosody learned during pre-training. In this way, effective pre-training with diverse text content can contribute to the performance of target speaker TTS system in the aspect of handling domain-mismatched text.

\vspace{-0.2em}
In some cases, it may not be convenient to collect extensive speech data with diverse content for model pre-training. Traditionally large amounts of text-audio parallel data could be acquired by: (1) studio recording with the target speaker(s) reading text scripts for many hours; or (2) downloading long speech recordings from the Internet, dividing them into sentence-level utterances and aligning with given text transcriptions \cite{chu2002domain, panayotov2015librispeech, zen2019libritts}. Both approaches are tedious, labour-intensive and costly. Using an excessive amount of pre-training data might also cause other concerns, namely long training time and high computational resources. For applications of speech generation in specific text domains, we investigate different approaches to reducing the required amount of pre-training data while maintaining desired TTS performance.

\vspace{-0.2em}
The contributions of this paper are as follows. First, we show that using diverse speech data to pre-train a speaker-independent TTS model can improve the performance of the target speaker TTS on domain-mismatched text. Second, test sets with different degrees of similarities to the text domain of target speaker data are designed.  The designed test sets are used to study how the pre-trained speaker-independent TTS model improves target speaker TTS performance; Third, We propose a method to improve the data and computational efficiency by reducing the pre-training data for a specific new text domain.

\begin{figure}[t]
  \centering
  \includegraphics[width=0.7\linewidth]{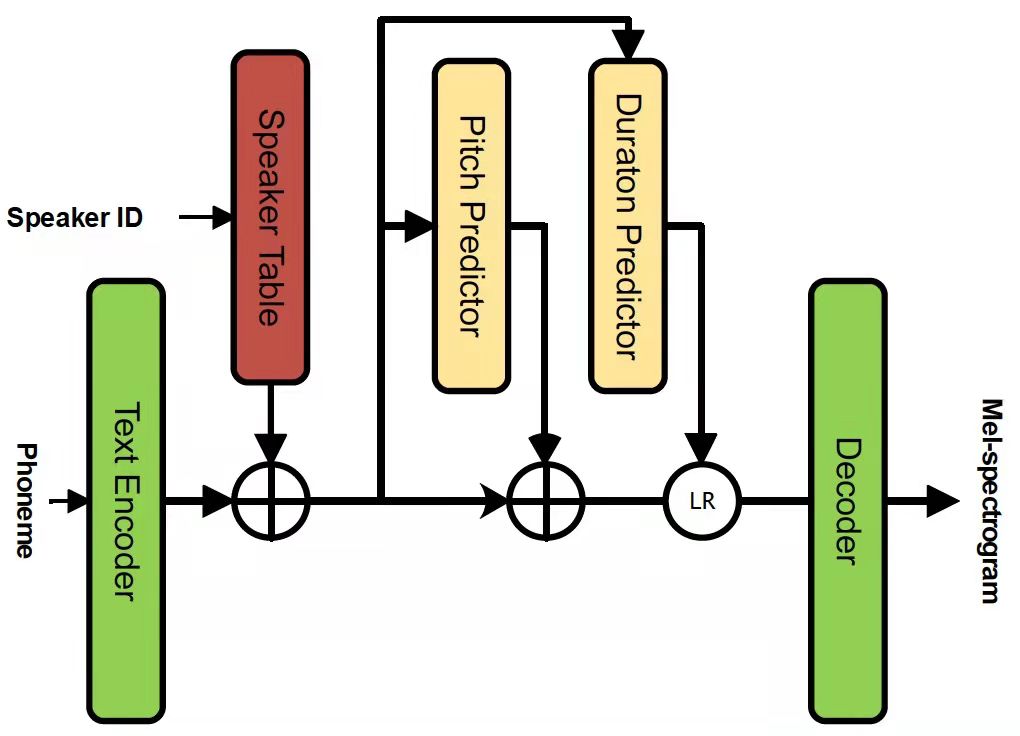}
  \caption{Diagram of TTS model}
  \label{fig:msfs2_pp}
\end{figure}

\vspace{-1.5em}
\section{Data Description}

\subsection{Pre-training Data}
Pre-training data are used to train a speaker-independent model, which is then fine-tuned on target speaker data. As our research questions concern the content coverage and domain of the pre-training data, a large-scale English speech corpus created for automatic speech recognition (ASR), namely LibriSpeech, is adopted \cite{panayotov2015librispeech}. LibriSpeech contains around $1000$ hours of speech data from $2484$ speakers. The speech content are based on the LibriVox’s audiobooks, which have a comprehensive coverage of topics. The complete LibriSpeech corpus is used as pre-training data. For the investigation on pre-train data reduction, 40,000 text-audio pairs (around 1/8 of the data in LibriSpeech) are used.

\vspace{-1em}
\subsection{Target Speaker Data}
The target speaker data for fine-tuning of the pre-trained speaker-independent TTS model are obtained from the LJSpeech corpus, which contains $13,100$ (about $24$ hours) audio clips of high-quality English speech produced by a female speaker reading passages from 7 non-fiction books. Experiments will be carried out to evaluate the performance of TTS systems fine-tuned with the whole LJSpeech corpus or its subsets of smaller size.

\vspace{-1em}
\section{Effect of Pre-training on target speaker TTS performance}
\label{epdmd}
Two TTS models are used to show how the pre-trained speaker-independent TTS model improves target speaker TTS performance. The first system is trained only on target speaker data, i.e., without pre-training with other data. This system is named as \textit{TTS w/o pre-training}. The second system, named \textit{TTS w/ pre-training}, is built by applying target speaker data (LJSpeech) to fine-tune a speaker-independent model pre-trained with LibriSpeech. The two systems are evaluated on three specially designed test sets, which have different degrees of similarity to the text content of target speaker data. The three test sets are named as \textbf{T-SIM}, \textbf{T-RAN}, \textbf{T-DIFF}, in the order from highest to lowest degree of similarity to the text content of target speaker data. 

\vspace{-0.7em}
\subsection{TTS Model Description}
The TTS model as shown in \autoref{fig:msfs2_pp} consists of a trainable speaker table and a speaker-independent TTS model. The speaker table models the voice characteristics for individual speakers. The speaker-independent model follows the non-autoregressive TTS system FastSpeech 2\cite{ren2020fastspeech}. In this study, the TTS model predicts pitch at phoneme level, which is found to give better TTS performance than frame-level prediction in FastSpeech 2. The text encoder converts the phoneme sequence into text embeddings, combined with speaker embeddings to predict the pitch and duration using the respective predictor modules. The length regulator (LR) module up-samples all phoneme-level features to frame-level features, which are then transformed into mel-spectrogram by the decoder. The Melgan vocoder\cite{kumar2019melgan} is used to generate speech waveform from the mel-spectrogram. 

\vspace{-1.1em}
\subsection{Design of Domain-mismatched Test Sets}
\vspace{-0.7em}
\label{test_set_prep}
We collect around 50,000 sentences from various sources that act as general text set to show the TTS performance on domain-mismatched text. Three test sets, each has $60$ sentences, are chosen from the general text set for subjective evaluation. A phoneme-based subword bigram language model is trained on the text domain of the target speaker data (LJSpeech).  The phoneme-level subwords for each word are obtained by applying Byte Pair Encoding (BPE)\cite{sennrich2016neural} to the phoneme sequence of the word. The vocabulary size of subwords is 200. The language model is applied to calculate the perplexity scores for all 50,000 sentences in the general text set. A sentence similar to the text domain of target speaker data will achieve low perplexity with the language model trained on that text domain. The text with the top 60 lowest perplexity scores make up the test set \textbf{T-SIM}, which is considered to lie in the same or similar domain of target speaker text data. The test set \textbf{T-DIFF}, which has a low degree of similarity to the text domain of target speaker data, comprises text with the top $60$ highest perplexity scores. The $60$ text in subset \textbf{T-RAN} are randomly sampled from the general text set.

\vspace{-1em}
\section{Pre-training Data Reduction for a new text domain}
\vspace{-1em}
\label{pdrda}
In the previous section, we aim to investigate how the pre-trained speaker-independent TTS model improves the target TTS system on the general text set. While in some cases, we want to improve the target speaker TTS performance on a specific new text domain. We assume that the speaker-independent model pre-trained on the data similar to the new text domain can effectively transfer the text-based prosody to the target speaker TTS system. To show this idea, around 9,000 text sentences from the novel books are prepared as the new text domain, which differs from the text domain of target speaker data. We aim to select a subset of LibriSpeech as pre-training data whilst achieving comparable performance for the target speaker TTS system.  Two methods are developed to select data with high text similarity to the new text domain from the LibriSpeech corpus. 

\vspace{-1.2em}
\subsection{Perplexity-Based Method}
\vspace{-0.7em}
\label{lm_pdr}
In this study, a phoneme-based subword bigram language model is trained on the new text domain. The text in the LibriSpeech corpus with low perplexities on the language model is assumed to have a high degree of similarity to the new text domain. Then, the corresponding pairs data in LibriSpeech will be selected as pre-training data.
  
\vspace{-1em}
\subsection{BERT-Based Method}
\label{bert_data_reduction}
\vspace{-0.7em}
BERT\cite{devlin2019bert} is a method that learns general-purpose text representation, which is trained on a large amount of open-domain text data.  In this study,  a pre-trained BERT model takes the text input to generate token representations for each sentence. The sentence-level vector is obtained by performing average pooling on token representations from the last encoder layer of BERT over a sentence. The new text domain can then be represented by the centroid of the sentence-level vectors belonging to the new text domain. The degree of similarity between each text and the new text domain is  measured by the L2 distance between the sentence-level vector of that text and the centroid vector of the new text domain. The data with high text similarities to the new text domain will be selected as pre-training data.

\vspace{-1em}
\section{Results and Analysis}
\vspace{-1em}
% \subsection{Subjective Evaluation Configurations}
All the subjective tests are evaluated via our internal crowdsourced listening test platform, with at least $15$ native judges for Comparison MOS (as CMOS) \cite{loizou2011speech} and $15$ judges for five-score MOS test for each test case.  Readers are recommended to listen to demo examples \footnote{https://patrick-g-zhang.github.io/pt-reduction/}. %The MOS results for Recording and Melgan copy synthesis are shown in \autoref{gt}.

% \begin{table}[htbp]
%         \caption{MOS for Recording and Melgan copy resynthesis.}
%     \centering
%   \scalebox{0.87}{
%     \begin{tabular}{m{2.5cm}m{2cm}m{4cm}}
%          \toprule 
%              & Recording & Melgan copy resynthesis  \\
%          \midrule
%          MOS   & $4.11\pm0.08$ &  $4.07\pm0.07$ \\
%          \bottomrule
%     \end{tabular}
%     }
%     \label{gt}
% \end{table}

\vspace{-1em}
\subsection{Subjective Evaluations on Domain-Mismatched Text}

\autoref{full_lj_comp} shows the MOS and CMOS comparison between the \textit{TTS w/o pre-training} and \textit{TTS w/ pre-training} on three test sets described in \autoref{test_set_prep}. The target speaker data is 24-hour LJSpeech, and the pre-training data is LibriSpeech. The \textit{TTS w/o pre-training} achieves the highest MOS score on \textbf{T-SIM} while the lowest on \textbf{T-DIFF}. This result indicates that the \textit{TTS w/o pre-training} performance will drop as the text is different from the text domain of target speaker data. On \textbf{T-DIFF} and \textbf{T-RAN}, the TTS performance of \textit{TTS w/ pre-training} is better than \textit{TTS w/o pre-training} by a large margin, showing that the pre-trained speaker-independent model improves the performance of target speaker TTS on the domain-mismatched text. Moreover, the MOS gap and CMOS between \textit{TTS w/o pre-training} and \textit{TTS w/ pre-training} on \textbf{T-DIFF} are more prominent than \textbf{T-RAN}, which means the improvement is more significant as the test text is more different from the text domain of the target speaker data. Concerning \textbf{T-SIM}, the \textit{TTS w/ pre-training} shows no performance gain (CMOS $-0.033$) in contrast to \textit{TTS w/o pre-training}, which might result from the fact that \textbf{T-SIM} is similar to the text domain of target speaker data (LJSpeech), thus the text-based prosody learned from the pre-trained speaker-independent model does not benefit the \textit{TTS w/ pre-training} on \textbf{T-SIM}. Since \textbf{T-RAN} is randomly sampled and can represent the general text set, we can claim the \textit{TTS w/ pre-training} has superior performance to the \textit{TTS w/o pre-training} on the domain-mismatched text.

% \vspace{-1em}

\begin{table}[htbp]
        \caption{The MOS and CMOS comparison on three test sets when the target speaker data is 24-hour LJSpeech.}
    \centering
    \scalebox{0.9}{
    \begin{tabular}{m{1.2cm}m{3cm}m{2.8cm}m{0.9cm}}
         \toprule 
          Set  & \textit{TTS w/o pre-training} & \textit{TTS w/ pre-training} & CMOS  \\
         \midrule
         T-SIM  & $\mathbf{3.95\pm0.06}$ & $3.93\pm0.07$ & $-0.033$ \\
         T-DIFF  & $3.67\pm0.07$ & $\mathbf{3.79\pm0.06}$  & $+0.318$ \\
         T-RAN  & $3.88\pm0.08$ &  $\mathbf{3.98\pm0.07}$  &$+0.287$ \\
         \bottomrule
    \end{tabular}
    }
    \label{full_lj_comp}
\end{table}

% \vspace{-1em}
As shown in \autoref{prosody_table}, the CMOS results of two TTS systems, in terms of pronunciation and prosody, are compared on \textbf{T-RAN}. When comparing the two systems, the raters only focus on pronunciation or prosody (tone and rhythm) rather than the overall impression of the speech.  Compared with the baseline \textit{TTS w/o pre-training} system, the \textit{TTS w/ pre-training} mainly improves prosodic variation of speech instead of pronunciation. This result agrees with our assumption that the main reason for \textit{TTS w/ pre-training} achieves improvement on the domain-mismatched text is that the text-based prosody learnt from pre-training data can be transferred to the target speaker TTS system.

\vspace{-1em}

\begin{table}[htbp]
        \caption{The pronunciation and prosody CMOS comparison on \textbf{T-RAN}.}
    \centering
    \scalebox{0.9}{
    \begin{tabular}{m{4.5cm}m{2.5cm}}
         \toprule 
             Focus setting & CMOS  \\
         \midrule
           pronunciation & $-0.041$ \\
           tone and rhythm & $+0.157$ \\
        %   2h & pronunciation & $+0.106$ \\
        %   2h & tone and rhythm & $+0.240$ \\
        
         \bottomrule
    \end{tabular}
    }
    \label{prosody_table}
\end{table}

\vspace{-1em}

Two TTS systems are also compared when we use a 1.5-hour subset of the LJSpeech corpus for fine-tuning, shown in \autoref{1_16_lj_comp}.  On all three test sets, the \textit{TTS w/ pre-training} performs much better than the \textit{TTS w/o pre-training}. Although 1.5 hours of speech data is sufficient for TTS system to produce intelligible speech\cite{chung2019semi}, the speech generated from system \textit{TTS w/o pre-training} still sounds unstable and jittery due to model overfitting. The improvement tendency among the three subsets is the same as that in \autoref{full_lj_comp}, where the improvement is the most notable on \textbf{T-DIFF} and the least marked on \textbf{T-SIM}. The \textit{TTS w/ pre-training} performs better than \textit{TTS w/o pre-training}  on \textbf{T-SIM} in this case, different from the corresponding result in \autoref{full_lj_comp} where target speaker data is over 20 hours.  After listening to the audios, we found the improvement on \textbf{T-SIM} might result from that \textit{TTS w/ pre-training} has more stable voice, which means the \textit{TTS w/ pre-training} mitigates the overfitting problem when the target speaker data is limited.

\vspace{-1em}

\begin{table}[htbp]
        \caption{The MOS and CMOS comparison on three test sets when the target speaker data is 1.5-hour LJSpeech.  }
    \centering
    \scalebox{0.9}{
    \begin{tabular}{m{1.2cm}m{3cm}m{2.8cm}m{0.9cm}}
         \toprule 
            Set & \textit{TTS w/o pre-training} & \textit{TTS w/ pre-training} & CMOS  \\
         \midrule
         T-SIM & $3.72\pm0.07$ & $\mathbf{3.81\pm0.06}$ & $+0.394$ \\
         T-DIFF  & $3.34\pm0.08$ & $\mathbf{3.65\pm0.07}$  & $+0.958$ \\
         T-RAN  & $3.66\pm0.08$ &  $\mathbf{3.84\pm0.07}$  &$+0.591$ \\
         \bottomrule
    \end{tabular}
    }
    \label{1_16_lj_comp}
\end{table}

\vspace{-1em}

\subsection{TTS Performance Evaluation on Pre-training Data Reduction for a New Text Domain}
\vspace{-0.5em}
For pre-training data reduction task, $60$ sentences are randomly sampled from the new text domain for subjective evaluation. Four target speaker TTS systems are evaluated, which is shown in \autoref{text_based_cmp}. All four target speaker TTS systems are fine-tuned from pre-trained speaker-independent TTS models. The only difference between those systems is the pre-training data: i) \textit{Random}: $40,000$ pairs data randomly sampled from the LibriSpeech, which is used as the baseline system; ii) \textit{Full}: LibriSpeech, which serves as the topline system; iii) \textit{Perplexity-based}: $40,000$ pairs data sampled from LibriSpeech with the method described in \autoref{lm_pdr}; iv) \textit{BERT-based}: 40,000 pair data selected from LibriSpeech using the method described in \autoref{bert_data_reduction}.
\vspace{-1em}
\begin{table}[htbp]
        \caption{The TTS performance comparison among four target speaker TTS systems.}
    \centering
    \scalebox{0.9}{
    \begin{tabular}{m{3cm}m{1.7cm}m{1.2cm}m{1cm}m{1cm}}
         \toprule 
            Pre-training data & MOS & CMOS1 & CMOS2\\
         \midrule
         \textit{Random}  & $3.84\pm0.08$ & $0.0$ & -\\
         \textit{Full}  & $\mathbf{3.94\pm0.07}$ & $+0.304$  & $0.0$\\
         \textit{Perplexity-based}  & $3.90\pm0.08$ &  $+0.246$  &$-0.063$\\
         \textit{BERT-based}  & $\mathbf{3.91\pm0.08}$ &  $+0.26$  & $-0.035$\\
         \bottomrule
    \end{tabular}
    }
    \label{text_based_cmp}
\end{table}

\vspace{-0.5em}

We conduct CMOS tests to compare the TTS performance between the two target speaker TTS model with pre-training data selected by proposed methods and target speaker TTS model with \textit{Random} pre-training data, shown in column CMOS1 in \autoref{text_based_cmp}.  It shows that the target speaker TTS systems with both \textit{BERT-based} and \textit{Perplexity-based} pre-training data can significantly improve TTS performance on the new text domain compared with the baseline system (target speaker TTS with \textit{Random} pre-training data). This confirms the effectiveness of the proposed pre-training data reduction approaches. The CMOS tests are also performed to compare the systems with the topline system, as shown in column CMOS2 in \autoref{text_based_cmp}.  It shows that target speaker TTS systems with both \textit{BERT-based} and \textit{Perplexity-based} pre-training data are  comparable to the topline system. %Two pre-training data reduction methods are compared too. The CMOS $+0.051$  indicates that the BERT-based is slightly better than the Perplexity-based pre-training data reduction method. 

\vspace{-1em}
\section{Conclusion}
\vspace{-1em}
In this study, we show that a pre-trained speaker-independent TTS model can improve the performance of the target speaker TTS model on domain-mismatched text compared with the TTS model trained only on the target speaker data. The subjective evaluation shows that the improvement is mainly on the prosody side instead of pronunciation with the help of the abundant text domains in pre-training. In order to improve the data efficiency of pre-training, two methods are applied to reduce the pre-training data of the speaker-independent TTS model for a new text domain. The target speaker TTS system with a selected subset of LibriSpeech as pre-training data can perform comparably to the target speaker TTS system with complete LibriSpeech as pre-training data, which significantly reduces the huge cost in pre-training data preparation.
\vfill\pagebreak
\label{sec:refs}

\bibliographystyle{IEEEbib}
\bibliography{strings,refs}

\end{document}